\documentclass{article}

\usepackage{microtype}
\usepackage{graphicx}
\usepackage{subfigure}
\usepackage{booktabs} 

\usepackage{hyperref}

\usepackage{amsmath}
\usepackage{amsfonts}
\usepackage{amssymb}
\usepackage{amsbsy}
\usepackage{amsthm}

\usepackage[accepted]{icml2018}

\icmltitlerunning{Thompson Sampling is differential private}

\newcommand{\Gauss}{\mathcal{N}}
\newtheorem{theorem}{Theorem}

\begin{document}

\renewcommand{\O}{\mathcal{O}}

\twocolumn[
\icmltitle{On The Differential Privacy of Thompson Sampling With Gaussian Prior}




\begin{icmlauthorlist}
\icmlauthor{Aristide C. Y. Tossou}{to}
\icmlauthor{Christos Dimitrakakis}{to}
\end{icmlauthorlist}

\icmlaffiliation{to}{Chalmers University of Technology, Sweden}

\icmlcorrespondingauthor{Aristide Tossou}{aristide@chalmers.se}
\icmlcorrespondingauthor{Christos Dimitrakakis}{chrdimi@chalmers.se}

\icmlkeywords{Multi-Armed bandits, differential privacy, thompson sampling}

\vskip 0.3in
]



\printAffiliationsAndNotice{}  



\section*{Summary of main results}
We show that Thompson Sampling with Gaussian Prior as detailed by Algorithm 2 in \cite{agrawal2013further} is already differentially private. Theorem \ref{theo:privacy_ts} show that it enjoys a very competitive privacy loss of only $\O(\ln^2 T)$ after T rounds. Finally, Theorem \ref{theo:regret_ts} show that one can control the privacy loss to any desirable $\epsilon$ level by appropriately increasing the variance of the samples from the Gaussian posterior. And this increases the regret only by a term of $\O(\frac{\ln^2 T}{\epsilon})$. This compares favorably to the previous result for Thompson Sampling in the literature (\cite{mishra2015nearly}) which adds a term of  $\O(\frac{K \ln^3 T}{\epsilon^2})$ to the regret in order to achieve the same privacy level. Furthermore, our result use the basic Thompson Sampling with few modifications whereas the result of \cite{mishra2015nearly} required sophisticated constructions.

\begin{theorem}[Differential privacy of Thompson Sampling with Gaussian Prior]
\label{theo:privacy_ts}
Thompson sampling with Gaussian Prior [Algorithm 2 in \cite{agrawal2013further}] is $\O(\epsilon, \delta)$-differentially private after $T$ rounds with $\epsilon = \ln^2 T$ and $\delta = T^{-4}$.
\end{theorem}

\begin{theorem}[Controllable privacy loss and regret of Thompson Sampling]
\label{theo:regret_ts}
If we run Thompson sampling with Gaussian Prior [Algorithm 2 in \cite{agrawal2013further}] and use for the variance of the Gaussian samples $\frac{\ln^2 T}{\epsilon(k_i +1)}$ then, we have an $\O(\epsilon, \delta)$-differentially private after $T$ rounds with $\delta = T^{-4}$. Furthermore, the expected regret of this algorithm is $\frac{\ln^2 T}{\epsilon} + \sqrt{K T \ln K}$.
\end{theorem}

\section*{Introduction}

The multi-armed bandit is a repeated
game with a fixed and finite set of $K$ actions between a player and an adversary. At each round $t$,
the player picks an action and simultaneously, the adversary
defines a reward for each action. The player will then receive the reward associated to the action he picked and is unaware of the reward of the remaining actions. When the reward are generated independently from a fixed
probability distribution given the action, we say that we are in the stochastic settings; otherwise, we are in the adversarial settings. The objective of the player is to get as much reward as he can and his performance (called \emph{regret}) is measured by how close he is to a benchmark policy. 

The multi-armed bandit is a model of many real-world applications and this has led multiple authors to consider the differential privacy guarantees one can achieve in this model. Most work has focused on the stochastic settings. \citet{mishra2015nearly},\cite{aaai16} provide differential private variants of a standard Upper Confidence Bounds Algorithm (UCB) \cite{auer2002finite} whereby the authors try to add carefully constructed noise to the original UCB. In our work, we also focus on the stochastic settings but provide guarantees for a different style of algorithm called Thompson Sampling. As opposed to UCB, Thompson Sampling is a randomized algorithm that works in a Bayesian framework. Indeed, one simply assigns a prior distribution to each action and at each round the player play the best action according the a sample from the posterior distribution. If one use the Gaussian Prior, the samples are generated from the Gaussian distribution with mean $\hat{u}_i$ (empirical mean) and variance $\frac{1}{k_i + 1}$ \cite{agrawal2013further}.
\citet{mishra2015nearly} provides some privacy guarantees for Thompson Sampling. However, their result require a complex modification of Thompson Sampling and include additional noise to the empirical mean of each action. This is in contrast to our work, whereby you do not add any external noise.

There have also been a few work \cite{tossou2017achieving}, \cite{agarwal2017price}, \cite{thakurta2013nearly} in the adversarial settings. However, this is not the object of our paper.

 
 In this work, we used the definition of differential private bandits as explained in Definition 2.2 of  \cite{tossou2017achieving}.

\section*{Results}

\begin{proof}[Sketch Proof of Theorem \ref{theo:privacy_ts}]
 According to \cite{kenthapadi2012privacy} (Lemma 1) A sample from $\Gauss(\mu, \sigma^2)$ is $(\epsilon_{k_i}, \delta)$-differentially private as far as $\sigma \geq \frac{\sqrt{2\ln(1/(2\delta)) + 2 \epsilon_{k_i}}}{w \cdot\epsilon_{k_i}}$  and $\delta < 1/2$ with $w$ the maximum change in $\mu$ when one single reward is changed. In our case, $w = \frac{1}{k_i+1}$. Solving the inequality and taking the minimum $\epsilon_{k_i}$ that satisfies it leads to 
 
 $\epsilon_{k_i} = \frac{1}{\sqrt{k_i+1}} + \sqrt{\frac{1 +  2\ln (\frac{1}{2\delta})}{k_i+1}}$
 
 To get the privacy loss over $T$ times
 we apply the advanced composition theorem of \cite{kairouz2017composition} (Theorem 3.5). Which leads to 
 the desired result.
\end{proof}

\begin{proof}[Sketch Proof of Theorem \ref{theo:regret_ts}]
The proof of the privacy is similar to that of Theorem \ref{theo:privacy_ts}.  The key idea in the proof of the regret is to observe that the increase in the variance only increases the number of pull of  suboptimal actions by $\frac{\ln^2T}{\epsilon}$.
\end{proof}

\section*{Conclusion}
Our results show that the original Thompson Sampling with Gaussian Prior enjoys a very competitive privacy guarantee and one can simply control the privacy level by increasing the variance. This change in the variance leads to a  negligible impact on the performance of the algorithm. This means that we are able to get privacy without any additional external noise or any complex tricks. The results of this work makes us wonder if one could apply similar techniques to achieve privacy in more complex problems. For example, it is an interesting open challenges if one can design a Bayesian deep learning algorithm that can simultaneously achieves differential privacy through the noise inherent in the algorithm.

\bibliography{bib}
\bibliographystyle{icml2018}

\end{document}